Pressure-dependent Intermediate Magnetic Phase in Thin $Fe_3GeTe_2$ Flakes

Heshen Wang[+,1], Runzhang Xu[+,2], Cai Liu[1], Le Wang[1], Zhan Zhang[1], Huimin Su[1], Shanmin Wang[3], Yusheng Zhao[3], Zhaojun Liu[4], Dapeng Yu[1], Jia-Wei Mei*[1], Xiaolong Zou*[2], Jun-Feng Dai*[1]

1. Shenzhen Institute for Quantum Science and Engineering, Southern University of Science and Technology, Shenzhen 518055, P. R. China
2. Shenzhen Geim Graphene Center and Low-Dimensional Materials and Devices Laboratory, Tsinghua-Berkeley Shenzhen Institute, Tsinghua University, Shenzhen 518055, P. R. China
3. Department of Physics, Southern University of Science and Technology, Shenzhen 518055, P. R. China
4. Department of Electrical and Electronic Engineering, Southern University of Science and Technology, Shenzhen 518055, P. R. China

+ The authors contribute to this work equally.
* Corresponding authors:
daijf@sustech.edu.cn; xlzou@sz.tsinghua.edu.cn; meijw@sustech.edu.cn;

**Abstract:**

We investigated the evolution of ferromagnetism in layered $Fe_3GeTe_2$ flakes under different pressures and temperatures using in situ magnetic circular dichroism (MCD) spectroscopy. We found that the rectangle shape of hysteretic loop under an out-of-plane magnetic field sweep can sustain below 7 GPa. Above that pressure, an intermediate state appears at low temperature region signaled by an 8-shaped skew hysteretic loop. Meanwhile, the coercive field and Curie temperature decrease with increasing pressures, implying the decrease of the exchange interaction and the magneto-crystalline anisotropy under pressures. The intemediate phase has a labyrinthine domain structure, which is attributed to the increase of ratio of exchange interaction to magneto-crystalline anisotropy based on Jagla's theory. Moreover, our calculation results reveal a weak structural transition around 6 GPa, which leads to a drop of the magnetic momentum of Fe ions.

**Introduction:**

Research on magnetic materials constructs the foundation of magnetic disk storages and spintronic devices. Since the intrinsic ferromagnetism in two-dimensional van der Waals crystals was discovered in 2017[1], scientists paid much more attention to searching for novel 2D ferromagnets due to lots of advantages in fabricating nanoscale electronic devices, such as chemical stability[2], effective modulation of ferromagnetism by external means[3-5]. The most important point is that ferromagnetism can maintain with decreasing the thickness of materials down to monolayer. So it can improve the density of data storage notably. Until now, it has been demonstrated experimentally that ferromagnetism can remain in $Fe_3GeTe_2$ (FGT) and $CrI_3$ monolayers[6-8]. For 2D Ising ferromagnet in $CrI_3$, it exhibits a layer-dependent transition between ferromagnetism and anti-ferromagnetism due to interlayer coupling and relative sliding between adjacent layers[9,10], which can be further tuned by applying an out-of-plane electric field[11-13] and hydrostatic pressures in experiments[14,15].

Similar to CrI$_3$, ferromagnetism of FGT shows a strong dimensional effect, where the ferromagnetic transition temperature dramatically decreases as the thickness is reduced down to 6 layers. Moreover, strong magneto-crystalline anisotropy (A) due to spin-orbit coupling as well as inter- and intra-layer exchange interaction (J) dominate the out-of-plane ferromagnetism in different levels. In contrast, the metalicity of FGT at room temperature and ambient pressure implies the significant electron itinerancy in the system. While, the gate-tunable room-temperature ferromagnetism in FGT[6] also suggests the subtle competition between the electron localization and itinerancy[16] in this system. Actually, for FGT, two inequivalent Fe ions, FeI$^{3+}$ and FeII$^{2+}$ (Fig. 1a) in monolayer, contribute to both the itinerant electrons and local ferromagnetic moments[17]. The unique magnetic behaviors and their interaction with interlayer coupling as well as the response to external pressures are intriguing but rarely discussed. Only until recently, one study showed pressurized FGT displayed a critical modulation of the anomalous Hall effect, without visible structural phase transition under pressures[18]. The magnetic evolution for FGT under pressures hasn't been explored yet.

In the present work, by means of hydrostatic pressure technique, the inter- and intra-layer distance among atoms can be tuned effectively, which has a direct effect on J and A and further influences the out-of-plane ferromagnetism. The evolution of ferromagnetism was monitored using in-situ magnetic circular dichroism (MCD) spectroscopy at temperatures from 33 K to room temperature and pressures from 3.7 GPa to 12.2 GPa. We observe that the ferromagnetic state with rectangular hysteresis loop suddenly disappear above 7.3 GPa, leaving behind an intermediate state with an 8-shaped skew loop at 33 K. The Curie temperature ($T_c$) for two states also dramatically decreases as pressure increases, which is due to the reduction of the local moments on Fe ions as well as pressure-induced increase of the electronic itinerancy. The enlarged intermediate phase region is probably due to the increase of ratio of J to A under pressure although both of them decrease with increasing pressure. With the aid of DFT calculations, we show that the magnetic momentum of FeI$^{3+}$ and FeII$^{2+}$ has a drop around 5-6 GPa due to the change of intralayer lattice structure, which may have the influence on the transition between ferromagnetic and intermediate states at low temperature.

**Main text**

High-quality FGT crystals were synthesized by the chemical vapor transport method [see Methods for details]. The X-ray diffraction (XRD) results, as shown in Fig. S1, confirmed the high structural quality of FGT crystals. Temperature-dependent magnetization of the bulk FGT, which was measured using a vibrating sample magnetometer (VSM) at field cooling mode with out-of-plane H, shows a transition from paramagnetic to ferromagnetic states around 200 K, as shown in Fig. S2. As reported in literature[6], the ferromagnetism of FGT mainly is derived from two inequivalent Fe ion sites, $FeI^{3+}$ and $FeII^{2+}$ in monolayer FGT (Fig. 1a). Meanwhile, partially filled d-orbitals of Fe atom dominate the energy band structure around the Fermi level, and lead to both electronic itinerancy and local ferromagnetic moments in bulk FGT. Moreover, strong intrinsic magnetocrystalline anisotropy due to spin-orbit coupling suppresses thermal fluctuations and thus preserves the 2D long-range ferromagnetic order.

Here, we employ diamond anvil cell (DAC) to achieve a high-pressure environment in layered FGT materials. To fabricate samples suitable for high-pressure experiments, the thin FGT films (around 20-50 nm) were mechanically exfoliated from the crystal using PC films and were deposited onto diamond culet by the dry transfer technique. The size of our experimental sample is around 50 μm × 50 μm, as shown in Fig. 1b. A rather large ratio of surface to volume guarantees that pressure is along c axis. The entire sample was covered by a thick hBN to prevent the sample degradation induced by the hydrostatic pressure medium. The bottom left image of Fig. 1b shows the optical image of our experimental sample, where the dashed line locates the thin FGT sample. After filling the pressure-transmitting medium and ruby balls, the entire DAC is mounted onto the cold finger of low-temperature cryostat with lowest temperature of 33K. The actual pressure inside cell is further determined by the peak energy of photoluminescence of ruby balls at low temperature (see Methods for details). Figure 1b shows a schematic of the magnetic circular dichroism (MCD) microscopy for an in-situ magnetic measurement under hydrostatic pressures. This method enables us to determine the coercive field and extract the ferromagnetic transition temperature under different pressures, and elucidates the evolution of the magnetic order with external pressures. All MCD measurements were performed at temperatures ranging from 33 K to 300 K in an out-of-plane magnetic field (see Methods for details).

We first present results from a thin FGT sample at a temperature of 33 K under different hydrostatic pressures, which are 3.7 GPa, 7.3 GPa and 10.8 GPa in first run and 4.3 GPa, 5.6 GPa, 9.6 GPa and 12.2 GPa in the second run. The two runs of pressurization experiments can ensure the reliability of experimental results. As revealed by in situ high-pressure synchrotron angle-dispersive XRD in ref. [18], the structural stability of FGT can be retained within the pressure of 25 GPa. Therefore, in our experimental condition, the highest pressure at 12 GPa cannot lead to any structural destruction. The main change is the shrinkage of lattice along c axis under pressure due to layered structure coupled by van der Waals force (Fig. S3). As shown in Fig. 2a, the FGT thin flake shows a typical rectangular hysteresis loop with a coercive field of 0.15 T at 3.7 GPa. It indicates a single domain exists in our experimental sample at low temperature. Below 7.3 GPa, the ferromagnetism is still observed as justified by the clear rectangular hysteresis loops at 4.3 GPa, 5.6 GPa and 7.3 GPa, respectively. However, the coercive field gradually decreases with increasing pressure from 0.15 T at 3.7 GPa to 0.087 T at 7.3 GPa, as shown in

Fig. 2b. For the magnetic sweeping in an out-of-plane mode, the square hysteresis loop indicates the easy magnetizing axis is along the out-of-plane direction. In the Stoner-Wohlfarth model for a single-domain[19], the coercive field $H_c$ is consistent with the magnetic anisotropy field $H_{ani}$, which is parallel with easy magnetizing axis. And its magnitude is equal to $2K\cos\gamma/M_s$, where K is the magnetic anisotropy energy density and $M_s$ is the saturation magnetization per volume. Based on this simple equation, we could infer that magnetic anisotropy energy density K decreases as pressure increases, so as to the magneto-crystalline anisotropy (A).

Above 7.3 GPa, FGT thin flake exhibits a completely different state, which is neither a ferromagnetic nor paramagnetic state. At high field (around ±0.2 T), the MCD signal approaches saturation value with opposite sign. When the field is swept from positive to negative magnetic field (B), the MCD signal suddenly jumps from the saturation value to an intermediate level in a positive $B_{in}$. Then decreases linearly as the magnetic field further changes. In the end, it approaches the reverse saturation, and vice versa. Actually, the magnetization jump at $|B_{in}|$ indicates a finite fraction of the spins change their orientations abruptly upon an infinitesimal range of the external field. This corresponds to a case where there is an abrupt nucleation, leading to a partially inverted magnetization state.

Actually, if we analyze the hysteresis loop at temperatures of 177 K and 186 K at 3.7 GPa, which is shown in Fig. 3a, the similar intermediate state can also be observed. This phenomenon has also been reported in Ref. [7] in thick FGT flakes at normal pressure at high temperature region. The magnetic force microscopy results showed the appearance of labyrinthine domain structure when the field is reduced through $|B_{in}|$, which is consistent with the image of abrupt nucleation in the sample. In comparison, for the pressure-related ferromagnetism in FGT in our study, we find that the intermediate state can be drove to low temperature region by applying the pressures. In the labyrinthine phase, the magnetization is easily tuned by the external magnetic fields and modulates the electric transport, which may have the potential application in the spintronics. We can expect that in an appropriate pressure, we can have the intermediate state even at zero temperature.

Fig. 3 shows MCD measurements as a function of magnetic field for several fixed temperatures at four representative pressures. At 3.7 GPa, the hysteresis loop gradually shrinks as temperature increases, and the intermediate state appears above $T_{c1}$=177 K, and then transforms to paramagnetic states above $T_{c2}$=203 K, as shown in Fig. 3a. It is consistent with the report in literature that FGT with thickness above 15 nm exhibits transition from single-domain ferromagnetic state to labyrinthine-domain ferromagnet to paramagnetic state as temperature increases. This kind magnetic transition among three states remains even at 7.3 GPa. However, the $T_{c1}$ and $T_{c2}$ are gradually decreases with increasing pressure. As shown in

Fig. 3a, they reduce to 96.5 K and 170 K at 7.3 GPa, respectively. At 10.8 GPa, we cannot observe obvious hysteresis loop at all the temperatures, but the transition from intermediate state to paramagnetic states still can be detected, with $T_{c2}$=122.5 at 10.8 GPa. With further increasing pressure, the $T_{c2}$ also further decreases as shown in Fig. 3b. The pressure-dependent phase diagram is summarized in Fig. 3c for three distinct magnetic states, namely single-domain ferromagnetic states, labyrinthine-domain ferromagnet and paramagnetic states. Below 7.3 GPa, three magnetic states are clearly observed. Above that pressure, ferromagnetic state suddenly disappears and leaves only the other states. $T_c$ is the critical temperature relating with the divergence of spin-spin correlation length. A decrease in $T_c$ implies a dramatic reduction in the energy scale of magnetic ordering in 2D FGT under pressure. In the mean-field theory, the Curie temperature is proportional to the exchange interaction J. According to the decreasing trend of $T_c$ with increasing pressure, we infer that the inter- and intra-layer exchange interaction will also decrease.

As reported in Jagla's theoretical paper[20], a scalar "phi-fourth" model was introduced to illustrate the change of hysteresis loop under an external z-oriented magnetic field. Three physical parameters, including magnetic exchange, anisotropy energy and structural disorder, control or influence the shape of loop. They found that hysteresis loop can be divided into two groups: weakly interacting ones for low J/A, and strongly interacting ones for large J/A. For the latter case, their model gives an anomalous 8-shaped hysteresis loop, where there is an abrupt nucleation, leading to a partially inverted magnetization state. This is consistent with the experimental results at the high pressure and low temperature (above 7 GPa). While the weakly interacting one shows the normal rectangle loop, which correspond to the case at the low pressure and low temperature. So we believe that the ratio of J/A is remarkably enhanced above 7 GPa, although both parameters decrease with increasing pressure.

We employed first-principles calculations to investigate structural, electronic, and magnetic properties of bulk $Fe_3GeTe_2$ under the application of hydrostatic pressure. After trying various pseudopotentials and DFT methods (see Methods for details) in calculating the magnetic properties, we found that the recently developed strongly constrained and appropriately normed (SCAN) method implemented in VASP code gave the closest agreement to the experimental observations with increasing hydrostatic pressure. At the ground state (without hydrostatic pressure), bulk $Fe_3GeTe_2$ is a ferromagnetic metal with magnetization easy axis along c-axis and magnetic moments of 2.71 $\mu_B$ and 1.80 $\mu_B$ on $FeI^{3+}$ and $FeII^{2+}$ ions, respectively. These values are larger than the reported experimental measurements[21-23] due to an overestimation of magnetic moment by DFT methods, as well as the underestimation of local moment in measurements of saturation moment.[21] After applying hydrostatic pressure, they are slightly weakened (~ -0.014 $\mu_B$/GPa) with increasing pressure (Fig. 4a), suggesting that the pressure has little effect on the occupancy of electronic orbitals and the weakened moment may be attributed to the pressure-induced structural change. Crossing the critical pressure around 5-6 GPa, there is a significant change in the Fe1-Te bond length (Fig. 4b). As a consequence, the magnetic moments of FeI and FeII have a drop. So we can expect that the structure transition would lead to a magnetism degradation. In the theoretical calculations, we find also an intralayer structure transition at critical pressure ($P_c$), i.e., the increase of FeI-FeI bond length, without significant changes in the lattice constant. This may explain that the structure transition in the previous pressured FGT is not observed in the XRD measurement[18]. Moreover, with increasing the pressure, the overlap of the Fe 3d electron wavefunctions increases (Fig. 4s), leading to the enhance of electron itinerancy[16], and a reduced magnetism in the materials as observed in the experiments.

**Conclusion**

In conclusion, we study the evolution of ferromagnetism of a layered FGT flake with pressures and temperatures using in-situ magnetic circular dichroism (MCD) spectroscopy. Our experimental results reveal that the coercive field and Curie temperature decrease as pressure increases and an intermediate state is drove to low temperature region above 7 GPa. We suggest that the increase of ratio of J to A contributes to the appearance of this state under low temperature. Moreover, DFT calculations show that the magnetic momentum of Fe ions decreases with increasing pressures, and has a drop around 5-6 GPa, which may have an influence on the transition between ferromagnetic state and intermediate state at low temperature region.

**Methods**

**Sample growth:** Single crystals of $Fe_3GeTe_2$ (FGT) were synthesized by Fe-Te flux as described in Ref [24]. The obtained layered crystals are soft and with metallic luster. By grounding single crystals into powder, the powder X-ray diffraction (XRD) pattern was obtained in Figure S1(a). The diffraction peaks are consistent with the Bragg reflections, suggesting the structure belongs to the space group P 63/m m c (No. 194). As shown in Fig. S1(b), the XRD from a flat layered crystal was indexed as (0 0 2) plane. The diffraction peaks marked by stars came from the residual Cu Kβ (λ = 1.39222 Å). No other impurity peaks are visible, revealing that $Fe_3GeTe_2$ single crystals is of high quality.

**High-Pressure Experiment:** The high-pressure were achieved using a diamond anvil cell (DAC) with maximum pressure up to 20 GPa. The diamond culets were polished along (100) crystal plane to minimize polarization-dependent response of the transmitted light. The chamber for loading samples and pressure medium is compose of a cylinder-shaped hole with a diameter of 300 μm and a height of 30 μm, which was fabricated in the center of a stainless-steel gasket. To obtain hydrostatic pressure conditions, the admixed methanol and ethanol in the molar ratio 4:1 was filled into the chamber to serve as the pressure-transmitting medium. A few ruby balls were also loaded into the chamber to indicate the internal pressure. Then, the face of two diamond-anvils was exactly aligned with each other, in order to achieve a uniform hydrostatic pressure on the sample. In the end, the entire DAC was mounted onto the cold finger of a closed-cycle cryostat (CRYO Industries) with the lowest temperature of 33 K and an out-of-plane magnetic field up to 2 T. The pressure in DAC was increased or decreased at room temperature, then was cooled down to the lowest temperature. The photoluminescence peak of ruby balls at low temperature was used to indicate the actual pressure.

**Magnetic circular dichroism measurements under high pressure**: For the in-situ magnetic circular dichroism (MCD) measurements, a 632.8 nm HeNe laser beam passed through a polarizer, a photoelastic modulator (PEM) and a beamsplitter, then was focused onto sample within DAV by a long working distance objective with spot size of around 2 μm. The reflected light went the same objective, beamsplitter and then was fed into a detector. The polarization state of incident light was modulated by the PEM, which periodically changes between linear and circular polarized light with frequency of 50 KHz. The corresponding AC signal that reflects the RMCD signal was collected by a lock-in amplifier.

**Density functional theory (DFT) based first-principles calculations:** The DFT calculations were carried out by using Viena ab-initio simulation package (VASP)[25,26] and a plane wave basis set.[27,28] The ion-electron interactions are treated by the projected-augmented wave (PAW) method,[27,28] while the exchange correlations are described by the spin-polarized Perdew-Burke-Ernzerhof (PBE) functional.[29] The spin-orbit coupling is considered to treat the heavy species and DFT-D3 method with Becke-Jonson damping[30,31] is used to treat the van der Waals interaction in all calculations. The bulk crystal structure was fully optimized with energy cut off set as 400 eV. The force convergence criteria used in relaxation was 0.01 eV/Å, while the energy convergence criteria were $10^{-5}$ eV for relaxation and further deceased to $10^{-7}$ eV for self-consistent calculations. The gamma-centered k-point mesh for structural relaxations and self-consistent calculations was set as 8×8×2 and 16×16×4 to ensure convergence. The recently developed strongly constrained and appropriately normed (SCAN) method[32,33] implemented in VASP code is used in self-consistent calculations of the magnetic properties of $Fe_3GeTe_2$ due to its closest agreement to experimental observations compared with other tested methods, including various pseudopotentials (PBE, PBE_sv_GW, LDA, and LDA_sv_GW) and DFT methods (GGA, GGA+U,[34] PBEsol,[35] and meta-GGA SCAN).

**Acknowledgements**
H.W. and R.X. contributed equally to this work. J.D. and J.M. conceived the project. J.D., H.W., Z.Z., H.S. and Z.L. designed and performed the experiments. C.L., L.W. and J.M. provided the experimental samples. S.W. and Y.Z. provided the DAV technique. R.X., X.Z. and J.M. provided the theoretical supports. All authors discussed the results and co-wrote the paper. The authors would like to thank Prof. Xiao-Dong Cui and Prof. Haizhou Lu for helpful discussions. J.F. acknowledges the support from the National Natural Science Foundation of China (11974159) and the Guangdong Natural Science Foundation (2017A030313023). J.W.M was partially supported by the program for Guangdong Introducing Innovative and Entrepreneurial Teams (No.~2017ZT07C062). Theoretical part was supported by the National Natural Science Foundation of China (Grant Nos. 11974197, and 51920105002), and Guangdong Innovative and Entrepreneurial Research Team Program (No. 2017ZT07C341).

**Conflict of Interest**
The authors declare no conflict of interest.

**Keywords**
2D ferromagnetic materials, magnetic circular dichroism (MCD) spectroscopy, hydrostatic pressure

Figures and captions

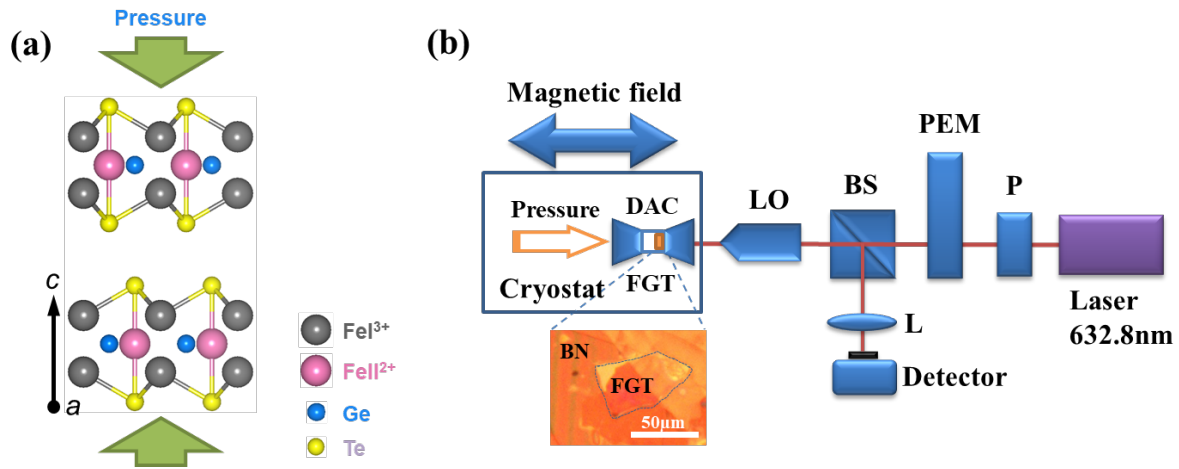

Figure 1: (a) Schematic of lattice structure of Fe$_3$GeTe$_2$ (FGT) under pressure. (b) Schematic of in-situ magnetic circular dichroism (MCD) experimental setup under hydrostatic pressure with the lowest temperature of 33 K and the highest magnetic field of 2 T. The thin FGT sample was covered with a thick hBN to avoid degradation induced by pressure media. P: Polarizer; PEM: photoelastic modulator; BS: beam splitter; DAC: diamond anvil cell; LO: 50× long-working distance objective; L: lens; FGT: thin Fe$_3$GeTe$_2$ sample.

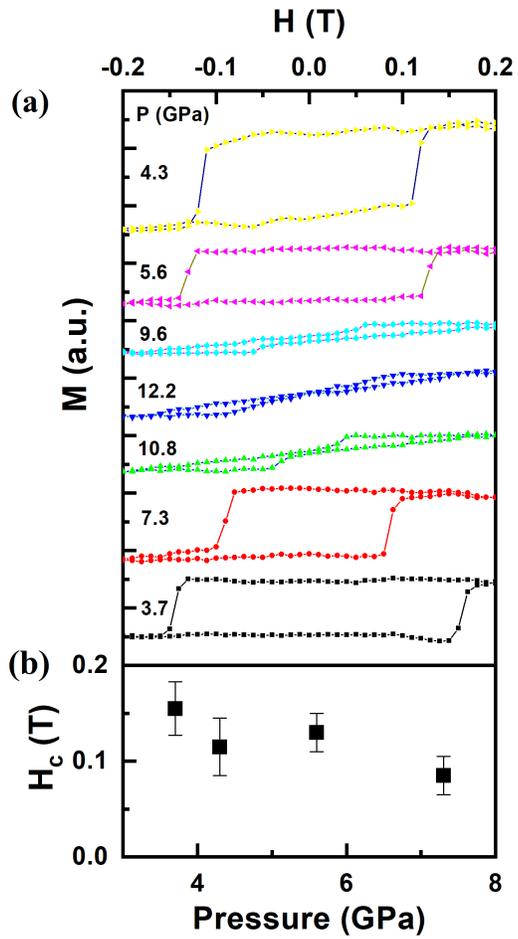

Figure 2: The ferromagnetic properties of thin FGT under pressure. (a) The magnetization (M) as a function of magnetic field (H) as various pressures, which are 3.7 GPa, 7.3 GPa and 10.8 GPa in the first run and 4.3 GPa, 5.6 GPa, 9.6 GPa and 12.2 GPa in the second run. (b) Extracted critical field for spin-flip transition at various pressures.

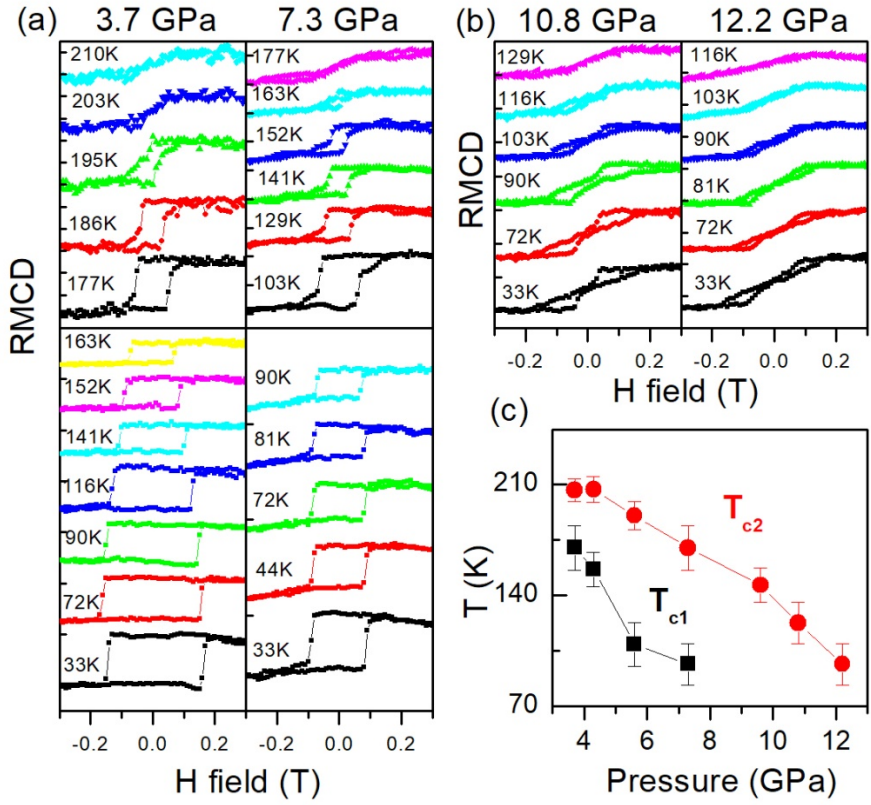

Figure 3: (a) Temperature-dependent MCD measurements as a function of magnetic field in the thin FGT flake at two representative pressures, namely 3.2 GPa and 7.3 GPa. Above 163 K at 3.7 GPa and 90 K at 7.3 GPa, as indicates by the line, intermediate magnetic states appear. Then FGT thin flake transforms to paramagnetic states above 203 K at 3.7 GPa and 163 K at 7.3 GPa, respectively. (b) MCD measurements at the pressures of 10.8 GPa and 12.2 GPa, respectively. Intermediate magnetic states appear at the lowest temperature of 33 K at both pressures. (c) The pressure-dependent phase diagram. Here $T_{c1}$ represents the transition temperature from ferromagnetic state to intermediate state, and $T_{c2}$ represents that from intermediate state to paramagnetic state.

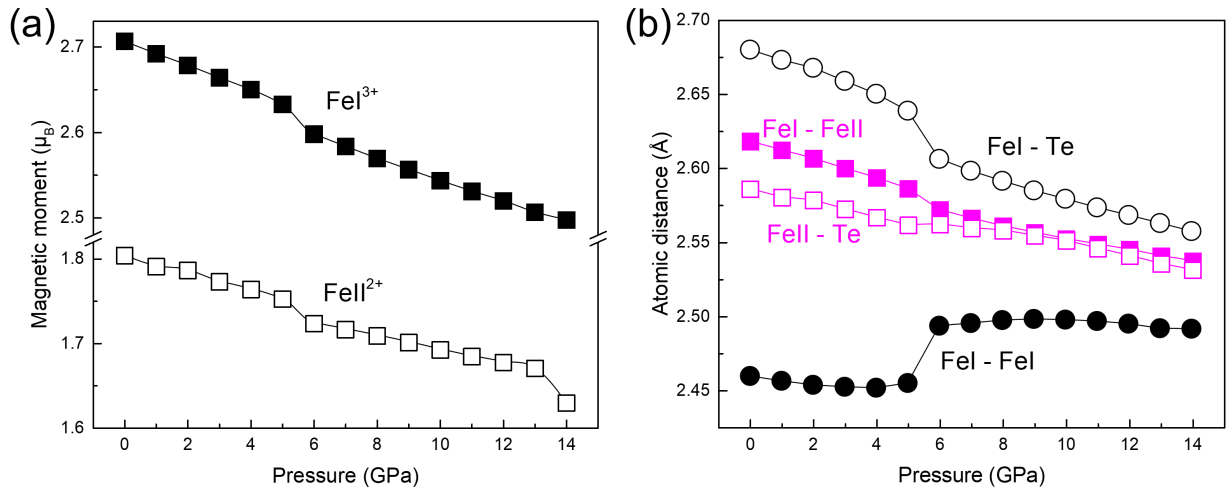

Figure 4: (a) Local magnetic moment of $FeI^{3+}$ and $FeII^{2+}$ ions as a function of hydrostatic pressure. Both moments decrease slightly in the range from 0 to 14 GPa. (b) Various ion-ion distances as functions of hydrostatic pressure.